\documentclass[12pt]{amsart}
\usepackage{amsmath}  

\newcommand{\cal}{\mathcal}
\newcommand{\fii}{\varphi}

\newcommand{\new}{\newcommand} 
\new{\bg}{\begin}
\newtheorem{lemma}{Lemma}
\newtheorem{definition}{Definition}
\newtheorem{proposition}{Proposition}

\newtheorem{remark}{Remark}

\new{\<}[1]{\begin{equation}{\label{#1}}} 
\new{\iii}{\begin{enumerate}} 
\new{\fff}{\end{enumerate}}
\new{\mfi}{\begin{eqnarray*}}             
\new{\mff}{\end{eqnarray*}}                 
\new{\mfni}{\begin{eqnarray}}             
\new{\mfnf}{\end{eqnarray}}                 
\new{\mai}[1]{\left(\begin{array}{#1}} 
\new{\maf}{\end{array}\right)}

\new{\scal}[2]{\mbox{$\langle{#1},{#2}\rangle$}}
\new{\ip}{\scal}
\new{\no}[1]{\left\|{#1}\right\|}
\new{\ket}[1]{|{#1}\rangle}
\new{\bra}[1]{\langle{#1}|}
\new\tr[1]{{\mathrm{tr}}\bigl[#1\bigr]}       
\new\hi{{\mathcal H}}
\new\ki{\mathcal K}
\new\room{\ \ \ \ \ \ }
\new{\toro}{\mbox{${\mathbf T}$}}
\new{\set}[1]{\{{#1}\}} 
\new{\runo}{\mathbb R}
\new{\eps}{\epsilon}
\new{\nor}[1]{\mbox{$\left\|{#1}\right\|$}}
\new{\cuno}{\mathbb C}
\new{\tuno}{\mathbb T}
\new{\hic}{\hi^{\chi}}
\new{\cuv}{c_{u,v}}
\new{\ocuv}{\overline{\cuv}}
\new{\cvu}{c_{v,u}}
\new{\ocvu}{\overline{\cvu}}
\new{\tuv}{T_{u,v}}
\new{\wpn}{W_{e_p}e_n}
\new{\mpp}{M_p(\pi)}
\new{\mpi}{M(\pi)}
\new{\tmpi}{\widetilde{M}(\pi)}
\new{\tf}{\tilde{f}}
\new{\tpi}{\tilde{\pi}}
\new{\tp}{\widetilde{P}}
\new{\lta}{L^2(X,\alpha)}
\new{\ltr}{L^2(\runo)}
\new{\lyr}{L^1(\runo)}
\new{\twu}{\widetilde{W}_u}
\new{\W}{B}

\begin{document}


\title[Phase space observables]{Phase space observables and isotypic spaces}
 
\author{G. Cassinelli}
\address{Gianni Cassinelli, Dipartimento di Fisica,
Universit\`a di Genova, I.N.F.N., Sezione di Genova, Via Dodecaneso~33,
16146 Genova, Italy}
\email{cassinelli@genova.infn.it}
\author{E. De Vito}
\address{Ernesto De Vito, Dipartimento di Matematica, Universit\`a di
Modena, via Campi 213/B, 41100 Modena, Italy and I.N.F.N., Sezione di Genova,
Via Dodecaneso~33, 16146 Genova, Italy}
\email{devito@unimo.it}
\author{P. Lahti} 
\address{Pekka Lahti, Department of Physics,
University of Turku, 20014 Turku, Finland}
\email{pekka.lahti@utu.fi}
\author{A. Levrero} 
\address{Alberto Levrero, Dipartimento di Fisica, Universit\`a di
Genova, I.N.F.N.,  Sezione di Genova, Via Dodecaneso~33, 16146 Genova,
Italy}
\email{levrero@genova.infn.it}
\begin{abstract}
We give  necessary and sufficient conditions for the set of Neumark
projections of a countable set of phase space observable to constitute 
a resolution of the identity, and we
give a criteria for a phase space observable to be informationally complete. 
The results will be applied to the phase space observables arising
from an irreducible representation of  the Heisenberg group.
\end{abstract}
\maketitle


\section{Introduction}\label{intro}
Phase space observables have turned out to be highly useful in various fields
of quantum physics, including quantum communication and information theory,
quantum tomography, quantum optics, and quantum measurement theory. Also many
conceptual problems, like the problem of joint measurability of noncommutative
quantities, or the problem of classical limit of quantum mechanics have
greatly advanced by this tool.  The
monographs~\cite{Davies,Helstrom,Holevo,Busch,Schroeck,Hakioglu,Perinova}
exhibit various aspects of these developments.

Any positive trace one operator $T$ (a state) defines a phase space observable
$Q_T$ according to the rule
$$
Q_T(E)= \frac 1{2\pi}\int_E e^{i(qP+pQ)}Te^{-i(qP+pQ)}dq\,dp, 
$$
where $E$ is a Borel subset of the (two dimensional) phase space.  It is
well known that all the phase space observables generated by {\it pure states}
have the same minimal Neumark dilation to a canonical projection measure on
$L^2(\runo^2)$.  On the other hand, the corresponding Neumark projections
depend on the pure state in question.  If $T$ is a pure state $|u\rangle\langle u|$ defined by a unit vector
$u$, we let $P_u$ denote the Neumark projection associated with $Q_{|u\rangle\langle u|}$ .
If two unit vectors  $u$ and $v$ are orthogonal then also $P_{u}P_{v}=0$.
One could then pose the problem of determining a set of orthonormal vectors 
$\{u_i\}$ such that the associated Neumark projections $\{P_{u_i}\}$ of the
phase space observables $Q_{|u_i\rangle\langle u_i|}$ constitute a resolution of the identity.
In~\cite{Lahti} it was shown that the set of number eigenvectors  possesses this
property.  This was proved by a direct method using the properties of the Laguerre
polynomials.

It turns out that this problem has a group theoretical background.
This follows from the work of A. Borel~\cite{Borel} on the group
representations that are square integrable modulo the centre.  Using
the results of Borel this problem can be traced back to the study of
the isotypic spaces of the representations induced by a central
character of the Heisenberg group $H^1$. (We recall that a
representation $(\pi,\hi)$ is called isotypic if it is the direct sum
of copies of the same irreducible representation).  More precisely, 
the phase space observables arise from an irreducible representation
of $H^1$ that is square integrable modulo the centre. This is actually
a general result: any irreducible representation $\pi$ of a group $G$
that is square integrable modulo the centre gives rise to covariant
``phase space observables'' with the above properties.  We prove that
a necessary and sufficient condition for the set of Neumark
projections $\{P_{u_i}\}$ to be a resolution of the identity is that
the representation of $G$ induced by the central character of $\pi$ be
isotypic.  This phenomenon occurs in particular for the Heisenberg
group, which is behind the phase space observables.

Phase space observables $Q_T$ that are generated by states $T$ such
that ${\rm tr\,}[Te^{i(qP+pQ)}]\ne 0$ for almost all
$(q,p)\in\runo^2$, are known to have another important property.
They are informationally complete, namely, if $W_1$ and $W_2$
are two states for which ${\rm tr\,}[W_1Q_T(E)]={\rm tr\,}[W_2Q_T(E)]$
for all $E$, then $W_1=W_2$, see, eg. \cite{AliPru, giape}.  We show
that, under suitable conditions, this property holds in general for
``phase space observables" associated with any irreducible
representations $\pi$ of $G$ square integrable modulo centre.

We hope that these results could bring further light on some of the many
applications of the phase space observables in quantum mechanics.

\section{Preliminaries and notations}\label{s1}
In this paper we use freely the basic concepts and results of harmonic
analysis, referring to~\cite{Folland95} as our standard source.  Let $G$ be a
Hausdorff, locally compact, second countable topological group, and let $Z$ be
its centre.  $Z$ is a closed, abelian, normal subgroup of $G$. We denote by
$X=G/Z$ the quotient space.  It is a Hausdorff, locally compact, second
countable topological group, and it is also a locally compact $G$-space with
respect to the natural action by left multiplication.  Let $p:G\to X$ be the
canonical projection and  $s:X\to G$  a Borel section for $p$, fixed
throughout the paper.

Assume further that $G$ is unimodular so that its left Haar measures
are also right Haar measures.  As an abelian subgroup $Z$ is also
unimodular.  We denote by $\mu$ and $\mu_0$ two (arbitrarily fixed)
Haar measures of $G$ and $Z$, respectively. Then there is a unique
$G$-invariant positive Borel measure $\alpha$ on $X$ such that for
each compactly supported continuous function $f\in C_c(G)$
\begin{equation}
\label{Weil}
\int_G\, f(g)\,d\mu(g) =\int_X\left(\int_Zf(s(x)h)\,d\mu_0(h)\right)\,
d\alpha(x).
\end{equation}
Moreover, $f\in L^1(\mu)$ if and only if the function 
$(x,h)\mapsto f(s(x)h)$ is in $L^1(\alpha\otimes\mu_0)$ and in this
case~(\ref{Weil}) holds for $f$.
The measure $\alpha$ is also a Haar measure for $X$ (regarded as a
group), both right and left.

We denote by $(\pi,\hi)$ a continuous unitary irreducible
representation of $G$ acting on a complex separable Hilbert space
$\hi$.  Let $h\in Z, g\in G$. Then
$\pi(h)\pi(g)=\pi(hg)=\pi(gh)=\pi(g)\pi(h)$, so that $\pi(h)$ commutes
with all $\pi(g), g\in G$. By Schur's lemma,
$$
\pi(h) = \chi(h)\, I,
$$
where $I$ is the identity operator on $\hi$ and $\chi$ is a
$\tuno$-valued character of $Z$, $\tuno$ denoting the group of complex
numbers of modulus one.  We call $\chi$ the central character of
$\pi$.

In the following we describe explicitly the imprimitivity system for
$G$, based on $X$, induced by the irreducible unitary representation
$\chi$ of $Z$.  There are several equivalent realizations of this
object, and we choose those which are most appropriate for our
purposes.

Let $\hic$ denote  the space of ($\mu$-equivalence classes of) 
measurable functions $f:G\to\cuno$ for which 
\iii
\item{}
$f(gh) = \chi(h^{-1})f(g)$ for all $h\in Z$,
\item{}
$f\circ s\in L^2(X,\alpha)$. 
\fff
The definition of the space $\hic$ does not depend on the
section $s$. Indeed, if $s'$ is another Borel section for $p$, then
for any  $x\in X$,
$s'(x)=s(x)h$ for some $h\in Z$, so that
$$
|f(s'(x))|^2=|f(s(x)h)|^2=|\chi(h^{-1})f(s(x))|^2 = |f(s(x))|^2.
$$
The space $\hic$ is a complex separable Hilbert space with respect to the
scalar product 
$$
\ip{f_1}{f_2}_{\hic} := \int_X \overline{f_1(s(x))}f_2(s(x))\,d\alpha(x),
$$
which is independent of $s$.

A description of the structure of $\hic$ is given by the
following property.  Let $K(G)^\chi$ denote the set of continuous
functions $f:G\to\cuno$ with the properties \iii
\item{}
$f(gh) = \chi(h^{-1})f(g)$ for all $g\in G, h\in Z$,
\item{}
$p({\rm supp}\, f)$ is compact in $X$.
\fff
If $\fii\in C_c(G)$, then the function $f_\fii$, defined by
$$
f_\fii(g) := \int_Z\chi(h)\fii(gh)\,d\mu_0(h),
$$
is in $K(G)^\chi$. Moreover, any function $f\in K(G)^\chi$ 
is of the form $f=f_\fii$ for some $\fii\in C_c(G)$ (see, e.g., \cite{Folland95}, Proposition 6.1, p. 152).
Obviously, $K(G)^\chi\subset\hic$ and $K(G)^\chi$ is dense in $\hic$.

The Hilbert space $\hic$ carries a continuous unitary representation $l$ of $G$
explicitly given by 
$$
(l(a)f)(g)  =  f(a^{-1}g),\ \ \ g\in G. 
$$
It is a 
realization of the representation of $G$ induced by the representation
$\chi$ of $Z$.

Let $\cal B(X)$ be the $\sigma$-algebra of the Borel subsets of $X$.
We define  a projection  measure on $(X,\cal B(X))$ by
$$
(P(E)f)(g):= \chi_E(p(g))f(g),
$$
where $E\in\cal B(X)$ and $f\in\hic$.  Clearly, $\cal B(X)\ni
E\mapsto P(E)\in B(\hic)$ is a projection measure and $(l,P)$ is an
imprimitivity system for $G$, based on $X$, and acting on $\hic$. Indeed,
$$
l(a)P(E)l(a)^{-1} = P(a.E),\ \ \ a\in G, E\in{\cal B}(X).
$$
It
is a realization of the imprimitivity system canonically induced by
$\chi$ and it is irreducible since $\chi$ is irreducible.

\section{Representations that are square integrable modulo the centre}
\label{s3}
Let $(\pi,\hi)$ be a continuous unitary representation of $G$ in a
complex separable Hilbert space $\hi$. Given $u,v\in\hi$, we denote by
$\cuv$ the function on $G$ defined through the formula
$$
\cuv(g) := \ip{\pi(g)u}{v}.
$$
This function  is called a {\em coefficient} of $\pi$ and
it is continuous and bounded,
$$
|\cuv(g)| = |\ip{\pi(g)u}{v}|\leq \no{\pi(g)u}\,\no{v} \leq
\no{u}\,\no{v}, \ \ \ g\in G,
$$
and it has the property $\cuv(gh) = \chi(h)^{-1}\cuv(g)$ for all $h\in Z$.
\begin{definition}
  {\rm Let $(\pi,\hi)$ be a continuous unitary irreducible
  representation of $G$.  We say that $\pi$ is {\em square
  integrable modulo the centre} of $G$, when, for all $u,v\in\hi$,
  $\cuv\circ s\in L^2(X,\alpha)$. }
\end{definition}

\noindent
This definition is independent of the choice of the function  $s$. Indeed, if
$s'$ is another section for $p$, then
$s'(x)=s(x)h$, $h\in Z$, for all $x\in X$, so that
$
\pi(s'(x)) = \pi(s(x)h)=\chi(h)\pi(s(x)),
$
and thus
$
|\ip{\pi(s'(x))u}{v}|^2=|\ip{\pi(s(x))u}{v}|^2.
$

We shall list next the basic properties of the square integrable
representations modulo the centre.  These results are due to A. Borel
\cite{Borel}, and they generalize the classical results of R. Godement
\cite{Godement} for square integrable representations.  \iii
\item{}
Let $\pi$ be a unitary irreducible representation of $G$ with central
character $\chi$. Then the following three statements are equivalent:
\newline
{}\ \ \ $a)$ $\pi$ is square integrable modulo $Z$;
\newline
{}\ \ \ $b)$ there exist $u,v\in\hi\setminus\{0\}$ such that 
$\cuv\circ s \in L^2(X,\alpha)$;
\newline
{}\ \ \ $c)$ $\pi$ is equivalent to a subrepresentation of $(l,\hic)$.
\item{}
If any (hence all) of the preceding conditions is satisfied, then $\cuv\in\hic$ for all
$u,v\in\hi$.
\item{}
If $(\pi,\hi)$ is square integrable modulo $Z$, there exists a number $d_\pi>0$, called
{\em the formal degree} of $\pi$, such that 
$$
\ip{\cuv}{c_{u',v'}}_{\hic}  =  
\frac 1{d_\pi}\ip{u'}{u}_\hi\,\ip{v}{v'}_\hi.
$$
The formal degree depends on the normalisation of the Haar measure $\mu$ so
that, possibily redefining $\mu$, one can always assume that $d_\pi=1$ so that
\begin{equation}\label{ort1}
\ip{\cuv}{c_{u',v'}}_{\hic}  =  \ip{u'}{u}_\hi\,\ip{v}{v'}_\hi.
\end{equation}
\item{} If $(\pi,\hi)$ and $(\pi',\hi')$ are two representations of
  $G$ which are square integrable modulo $Z$, whose central characters
  $\chi$ and $\chi'$ coincide, and which are not equivalent, then
\begin{equation}\label{ort2}
\ip{\cuv}{c'_{u',v'}}_{\hic} = 0,
\end{equation}
where $c'_{u',v'}$ are coefficients of $(\pi',\hi')$.
\fff

\section{Canonical POM associated with a square integrable representation
modulo the centre}\label{s4}
Let $(\pi,\hi)$ be a fixed representation with central character $\chi$ and 
square integrable modulo the centre.
Fix $u\in\hi\setminus\{0\}$, and define 
$
W_u:\hi\to\hic
$
by
$$
W_uv := \cuv,\ \ \ v\in\hi.
$$
 $W_u$ is a linear map  and it is a multiple of an isometry. Indeed, if $v,w\in\hi$, then
\begin{equation}\label{ort3}
\ip{W_uv}{W_uw}_{\hic} =
\no{u}^2_\hi\,\ip{v}{w}_\hi.
\end{equation}
The range of $W_u$ is a closed subspace of $\hic$, and
$1/\no{u}_\hi\, W_u$ is a unitary operator from $\hi$ to the range of $W_u$.
The operator $W_u$ intertwines the action of $\pi$ on $\hi$ with the action of $l$ on $\hic$. In fact,
for any $a\in G$,
\mfi
(W_u(\pi(a)v))(g) &=& c_{u,\pi(a)v}(g)=
\ip{\pi(g)u}{\pi(a)v}_\hi\\
&=& \ip{\pi(a^{-1}g)u}{v}_\hi = c_{u,v}(a^{-1}g)\\
&=& (W_uv)(a^{-1}g) = (l(a)(W_uv))(g),
\mff
showing that
$$
W_u\,\pi(a) = l(a)\, W_u
$$
for all $a\in G$. Hence ${\rm ran}\,W_u$ is invariant with respect to $l$ and
the unitary operator $1/\no{u}_\hi\, W_u$  defines an isomorphism of the unitary irreducible 
representations   $(\pi,\hi)$  and  $(l|_{\,{\rm ran}\,W_u}, {\rm ran}\,W_u)$ of $G$,
$$
(\pi,\hi) \simeq (l|_{\,{\rm ran}\,W_u}, {\rm ran}\,W_u).
$$
We are in a position to associate to any state $T$ 
a natural positive operator measure
(POM) on $(X,\cal B(X))$, with values in the positive operators on
$\hi$. Given a state $T$, for all $E\in\cal B(X)$ we define 
\begin{equation}\label{ronza}
Q_T(E) =  \int_E \pi(s(x)) T \pi(s(x))^{-1} \,d\alpha(x),
\end{equation}
where the integral is in the weak sense. The definition is well posed.
Indeed, let $T=\sum_i \lambda_i |e_i\rangle\langle e_i|$ be the spectral
decomposition of $T$ and fix a trace class operator $\W$ with the 
decomposition $\W=\sum_k w_k |u_k\rangle\langle v_k| $,
where $w_k\geq 0$ and  $(u_k), (v_k)\subset\hi$  are orthonormal sequences. 
Since $\pi$ is square integrable modulo $Z$, 
the function
$$ 
\phi_{ik}(x)=\overline{c_{e_i,v_k}(s(x))}c_{e_i,u_k}(s(x))
= \ip{v_k}{\pi(s(x))e_i}\ip{e_i}{\pi(s(x))^{-1}u_k}  
$$
is $\alpha$-integrable on $X$ and, using the H\"older  inequality and
the orthogonality relations~(\ref{ort1}), 
\mfi
\int_E |\phi_{ik}(x)|\,d\alpha(x) & \leq &
\left( \int_E |c_{e_i,v_k}(s(x))|^2\,d\alpha(x) \right)^{\frac 12} 
\times\\
& & \room \left( \int_E |c_{e_i,u_k}(x)|^2\,d\alpha(x) \right)^{\frac 12}\\
&\leq& \no{c_{e_i,v_k}}_{\hic}\nor{c_{e_i,u_k}}_{\hic} \\
& \leq & \nor{e_i}^2_\hi  \nor{v_k}_\hi \nor{u_k}_\hi  =  1.
\mff 
Since $\sum_{i,k}\lambda_i w_k =
\nor{T}_1\nor{\W}_1=\nor{\W}_1$, the series  $\sum_{i,k}\lambda_i w_k \phi_{ik}$
converges $\alpha$-almost everywhere to an integrable function $\phi$ and
$$
\int_E\phi(x)\,d\alpha(x)=\sum_{i,k}\lambda_i w_k \int_E\phi_{ik}\,d\alpha(x).
$$
On the other hand, for $\alpha$-almost all $x\in X$, $\phi(x)=\tr{\W\pi(s(x))T\pi(s(x))^{-1}}$.
Hence $\int_E |\tr{\W\pi(s(x))T\pi(s(x))^{-1}}|\,d\alpha(x) \leq 
\nor{\W}_1$
 and the linear form 
$$\W\mapsto \int_E \tr{\W\pi(s(x))T\pi(s(x))^{-1}}\,d\alpha(x)$$
is continuous
on the Banach space of the trace class operators. Therefore it defines a bounded
operator $Q_T(E)$ such that 
\mfi
\tr{\W Q_T(E)} & = & \int_E \tr{\W\pi(s(x)) T \pi(s(x))^{-1}}\,d\alpha(x) \\
& = & \sum_{i,k}\lambda_i w_k \int_E \ip{v_k}{\pi(s(x))e_i}\ip{e_i}{\pi(s(x))^{-1}u_k}\,d\alpha(x)\\
& = & \sum_{i,k}\lambda_i w_k \int_E \overline{c_{e_i,v_k}(s(x))}c_{e_i,u_k}(s(x)) \,d\alpha(x).
\mff
By choosing $\W=|u\rangle\langle v|$ we see that $Q_T(E)$ has the
expression~(\ref{ronza}).

The mapping $E\mapsto Q_T(E)$ defines a POM on $X$.
Indeed, $Q_T(E)$ is a positive operator and, given $u,v\in\hi$,
the map $E\mapsto\ip{u}{Q_T(E)v}_\hi$ is a complex 
measure on $(X,\cal B(X))$,
due to the $\sigma$-additivity  of the integral.

Moreover, $Q_T(X)=I$. Indeed,  for all $u,v\in\hi$,
\mfi
\ip{u}{Q_T(X)v}_\hi  
&=&\sum_i \lambda_i 
\int_X \overline{c_{e_i,u}(s(x))}c_{e_i,v}(s(x))  \, d\alpha(x) \\
&=& \sum_i\lambda_i\ip{c_{e_i,u}}{c_{e_i,v}}\\
&=& \sum_i \lambda_i  \nor{e_i}^2_\hi \ip{u}{v}_\hi
= \ip{u}{v}_\hi.
\mff

The operator measure $E\mapsto Q_T(E)$ is covariant under the representation $(\pi,\hi)$,
that is, for all $E\in\cal B(X), a\in G$,
$$
\pi(a)Q_T(E)\pi(a)^{-1} = Q_T(a. E).
$$
Indeed,
\mfi
\pi(a)Q_T(E)\pi(a)^{-1} 
& = &\int_E \pi(a)\pi(s(x)) T \pi(s(x))^{-1} \pi(a)^{-1} \,d\alpha(x) \\
& = &\int_E \pi(as(x)) T \pi(as(x))^{-1} \,d\alpha(x) \\
& = &\int_E \pi(s(a.x)) T \pi(s(a.x))^{-1} \,d\alpha(x) \\
& = &\int_{a.E} \pi(s(x)) T \pi(s(x))^{-1} \,d\alpha(x) \\
& = &Q_T(a. E)
\mff
where we used the fact that
$as(x)=s(a. x)h$, for some $h\in Z$.

\section{The minimal Neumark dilation of $Q_u$}
In this section we consider the operator measure $Q_{|u\rangle\langle u|}$ associated with
a pure state $|u\rangle\langle u|$ and we show that the canonical projection
measure $P$ defined in Section~\ref{s1} is the minimal Neumark dilation of 
$Q_{|u\rangle\langle u|}$ for any $u$.

Given a unit vector $u\in \hi$,  we denote simply by $Q_u$
the POM $Q_{|u\rangle\langle u|}$. 
Then for any  $E\in{\cal B}(X)$ and  for all $v,w\in\hi$, 
\mfi
\ip{W_uv}{P(E)W_uw}_{\hic} &=&
\int_X \overline{(W_uv)(s(x))}\,\chi_E(x)\, (W_uw)(s(x))\, d\alpha(x)\\
& = & \int_E \overline{c_{u,v}(s(x))} c_{u,w}(s(x))\, d\alpha(x) \\
&=&\ip{v}{Q_u(E)w}_\hi,
\mff
which shows that $P$ is a Neumark dilation of  $Q_u$.

Furthermore, $P$ is minimal in the sense that $\hic$ is the smallest closed
space containing all the vectors of the form $P(E)f$, as $E$ varies in $\cal
B(X)$ and $f$ varies in ${\rm ran}\, W_u$,
$$
\hic = \overline{{\rm span}}\,\{P(E)f\,|\, E\in\cal B(X), f\in\ {\rm ran}\, W_u\}.
$$
We go on to prove this fact. Due to the irreducibility of $\pi$, all the vectors of $\hi$ are
cyclic for $\pi$ itself. Hence for any $v\in\hi$, $v\neq 0$,
$$
\hi = \overline{{\rm span}}\,\{\pi(a)v\,|\, a\in G\}.
$$
Therefore,
$$
{\rm ran}\, W_u = \overline{{\rm span}}\,\{ W_u(\pi(a)v)\,|\, a\in G\}
=  \overline{{\rm span}}\,\{ (l(a)W_u)(v)\,|\, a\in G\},
$$
so that 
\mfi
&& \overline{{\rm span}}\,\{P(E)f\,|\, E\in\cal B(X), f\in\ {\rm ran}\, W_u\}\\
&=& \overline{{\rm span}}\,\{ P(E)(l(a)W_u)(v)\,|\,  E\in\cal B(X), a\in G\}\\
&=& \overline{{\rm span}}\,\{ l(a)P(a^{-1}. E)W_u(v)\,|\, E\in\cal
B(X), a\in G\}.  
\mff 
Now $W_uv$ is a nonzero element of $\hic$ and $(l,P)$ is an
irreducible imprimitivity system for $G$, acting in $\hic$, so that
$$
\overline{{\rm span}}\,\{ l(a)P(a^{-1}. E)W_u(v)\,|\,  E\in\cal B(X), a\in G\}=\hic,
$$
which completes the proof of the statement.

As a final remark we notice that the Neumark projection $P_u:\hic\to\hic$ onto the range of $W_u$ is explicitly given by $P_u=W_uW_u^*$.

\section{A decomposition of the space $\hic$}\label{s5}

In this section we describe a decomposition of the space $\hic$ associated with the representation
$(\pi,\hi)$ of $G$.
We denote
\mfi
M(\pi)_0 & := & \sum_{u\in\hi}{\rm ran\,}W_u \\
& = & {\rm span}\,\{\cuv\,|\, u,v\in\hi\}  \\
 M(\pi) & := & \overline{M(\pi)_0 }.
\mff
$M(\pi)$ is the smallest closed subspace of $\hic$
that contains all the ranges of the maps $W_u$.
If $\pi$ and $\pi'$ are equivalent representations, then
$$
M(\pi)=M(\pi').
$$
In  other words, $M(\pi)$ depends only on the equivalence class of $\pi$.
On the other hand, 
if $\pi$ and $\pi'$ are not equivalent, but they have the same central character $\chi$,
then the orthogonality condition~(\ref{ort2})  imply that
$$
M(\pi)\perp M(\pi').
$$
We proceed to study the structure of the subspace $M(\pi)$.
 
\

\iii

\item{}
 $M(\pi)$ is invariant under the action of $l$.
This is clear since $M(\pi)_0$ is invariant with respect to $l$, hence, for any 
$a\in G$, $l(a)M(\pi)=\overline{l(a)M(\pi)_0}=\overline{M(\pi)_0}=M(\pi)$.

\

\item{} Let $(e_n)_{n\geq 1}$ be a basis of $\hi$.  Then $(\wpn)_{n,p\geq 1}$
  is a basis of $M(\pi)$.  To show this, observe that
  $\ip{\wpn}{W_{e_q}e_m}_{\hic}= \ip{e_q}{e_p}_{\hi}\ip{e_n}{e_m}_{\hi} $, so
  that $(\wpn)_{n,p\geq 1}$ is an orthonormal set in $M(\pi)$.  Given
  $u,v\in\hi$, one has that
  $\sum_{n,p}|\ip{u}{e_n}\ip{e_p}{v}|^2=\nor{u}^2\nor{v}^2$.  Hence the series
  $\sum_{n,p}\ip{u}{e_n}\ip{e_p}{v}\wpn$ converges in $M(\pi)$. Since, for all
  $g\in G$, $\sum_{n,p}\ip{u}{e_n}\ip{e_p}{v}\wpn(g)$ converges to $W_uv(g)$,
  the set $(\wpn)_{n,p\geq 1}$ generates $M(\pi)_0$, hence $M(\pi)$.

\

\item{} 
  The space $M(\pi)$ is isotypic, in fact it can be decomposed as
  $$
  \mpi=\oplus_{p\geq 1}\,{\rm ran\,}W_{e_p}
  $$
  and, for any $p$ the representation $(l|_{{\rm ran\,}W_{e_p}},{\rm
  ran\,}W_{e_p})$ is unitarily equivalent to $(\pi,\hi)$.

\fff

\

The Hilbert sum of the subspaces $\mpi$, as $\pi$ runs through the
(inequivalent) irreducible representations of $G$ with the same
central character $\chi$ that are square integrable modulo the centre,
does not exhaust $\hic$, in general.  This Hilbert sum is the {\em
  discrete part} of $\hic$.  In fact, let $V$ be a closed subspace of
$\hic$ which is invariant and irreducible under $l$, and denote by $\sigma$ the
restriction of $l$ to $V$. Then $\sigma$ is a square integrable
representation of $G$ modulo the centre, with the same central
character $\chi$, and one has the following result.
\begin{proposition}\label{prop}
  The subspace $V$ is contained in $M(\sigma)$.
\end{proposition}
\begin{proof}
Let $f\in V$ and denote by $S:\hic\to V$ the orthogonal projection
onto $V$. For all $g\in\hic$ and $a\in G$ we have
$$
\ip{\sigma(a)Sg}{f}_{\hic}  =  \ip{Sl(a)g}{f}_{\hic} = \ip{l(a)g}{f}_{\hic}.
$$
Since $Sg$ and $f$ are in $V$ and $(\sigma,V)$ is square integrable
modulo $Z$, we have 
$$ 
\left( a\mapsto \ip{l(a)g}{f}_{\hic} \right)\in M(\sigma).
$$
Explicitly, 
$$
\ip{l(a)g}{f}_{\hic} = \int_X \overline{g(a^{-1}s(x))}
f(s(x))\,d\alpha(x).
$$
For any $\phi\in C_c(G)$ the function $f_\phi$ defined in
section~\ref{s1}
is in $K(G)^\chi\subset\hic$ and we get
$$
\ip{l(a)f_\phi}{f}_{\hic}= \int_X d\alpha(x) f(s(x)) \int_Z d\mu_0(h)
\overline{\chi(h)}\overline{\phi(a^{-1}s(x)h)}.
$$
We claim that
$$
\left(
x,h\mapsto f(s(x)) \overline{\chi(h)}\overline{\phi(a^{-1}s(x)h)}
\right)\in L^1(\alpha\otimes\mu_0).
$$
Indeed 
$$
\int_Z |f(s(x)) \chi(h)\phi(a^{-1}s(x)h)| \,d\mu_0(h)
= |f(s(x))|\int_Z |\phi(a^{-1}s(x)h)|\,d\mu_0(h)
$$
and the function 
$$ x\mapsto \int_Z |\phi(a^{-1}s(x)h)|\,d\mu_0(h) $$
is in $C_c(X)$ (see, for instance,~\cite{Folland95}).
Hence its product with $|f(s(x))|$ is in $L^1(\alpha)$ and the claim
follows by Tonelli's theorem.
Now we can apply Equation~(\ref{Weil}) to the function 
$$
f(s(x)) \overline{\chi(h)}\overline{\phi(a^{-1}s(x)h)}= 
f(s(x)h) \overline{\phi(a^{-1}s(x)h)}
$$
to conclude that 
\mfi
\ip{l(a)f_\phi}{f}_{\hic} & = &
\int_G f(g) \overline{\phi(a^{-1}g)}\,d\mu(g) \\
& = & (f*\tilde\phi)(a),
\mff
where $\tilde\phi(a):=\overline{\phi(a^{-1})}$, and $*$ denotes the convolution.
In particular $f*\tilde\phi\in M(\sigma)$.
If we let $\phi$ run over a 
sequence of functions on $G$ which is an approximate 
identity, see for example~\cite{Folland95},  one can prove that 
$f*\tilde\phi \to f$ in $\hic$ (see the below remark) 
and, since $M(\sigma)$ is closed, $f\in M(\sigma)$. 
This shows that $V\subseteq  M(\sigma)$.
\end{proof}
\begin{remark}
  {\rm The proof of the above proposition uses the fact that
    $f*\tilde\phi \to f$ in $\hic$ when $\phi$ runs over a sequence of
    functions on $G$ which is an approximate identity. To
    show this technical result one can mimic the standard proof in
    $L^2(G)$, taking into account that there exists a Borel measure
    $\nu$ on $G$ having density with respect to $\mu$ such that the
    induced representation $(l,\hic)$ can be realized on a suitable
    subspace of $L^2(G,\nu)$ (compare Ex.~6,~Sect~XXII.3
    of~\cite{dieu}).}
\end{remark}

To summarize,
$$\hic=\oplus_\pi M(\pi) \oplus R,$$
where the first direct sum ranges over the inequivalent irreducible
representations of $G$ with central character $\chi$ that are square
integrable modulo the centre and the orthogonal complement $R$ is the
continuous part of the decomposition.

We can now state the main result of the paper.
\begin{proposition} 
  Let $(\pi,\hi)$ be a square integrable representation of $G$ modulo
  the centre.  Let $\{e_i\}$ be a basis of $\hi$. Then the set of
  orthogonal projections $\{ W_{e_i}W_{e_i}^*\}$ is a resolution
  of the identity in $\hic$ if and only if $(l,\hic)$ is an isotypic
  representation.
\end{proposition}
\begin{proof}
From items 2 and 3  above it follows that  the set $\{ W_{e_i}W_{e_i}^*\}$ is a
resolution of the identity of $\mpi$ and $(l,\mpi)$ is an isotypic
representation. Hence, $\{ W_{e_i}W_{e_i}^*\}$ is a resolution of
the identity in $\hic$ if and only if $\mpi=\hic$ and, in this case,
$(l,\hic)$ is isotypic. Conversely, assume that
$(l,\hic)$ is an isotypic representation. Let $(\sigma,V)$ be an
irreducible subrepresentation of $(l,\hic)$, then $\sigma$ is square
integrable modulo the centre and, by Proposition~\ref{prop}, $V\subset
M(\sigma)$. Since $(l,\hic)$ is isotypic and $\pi$ is equivalent to a
subrepresentation of $(l,\hic)$, $\sigma$ is equivalent to $\pi$, so
that $M(\sigma)=\mpi$. Since $\hic$ is direct sum of copies of
$(\sigma,V)$, it follows that $\hic=\mpi$.
\end{proof}

\section{The informational completeness}
An interesting property of the phase space observables is related to the notion of 
informational completeness. We say that the operator measure $Q_T$, associated with the
state $T$, is informationally complete if the set of operators $\{Q_T(E)\,|\, E\in\cal B(X)\}$
separates the set of states, \cite{giape, Prugo}. An extensive study
of the conditions assuring the informational completeness is given
in~\cite{jmp}. In this section, we prove some results suited to our
case. First of all,
\begin{lemma}
Let $T$ be a state in $\hi$ and $Q_{T}$ the corresponding POM generated by
the representation $\pi$. Then the following conditions are equivalent:
\iii
\item $Q_{T}$ is informationally complete;
\item if $\W $ is a trace class operator and $\tr{\W \pi(g)T\pi(g^{-1})}=0$
  for all $g\in G$, then $\W =0$. 
\fff
\end{lemma}
\begin{proof}
It is known, see for example \cite{giape}, that $Q_T$ is informationally
complete if and only if it separates the set of trace class operators. Let $\W$
be a trace class  operator, then $\tr{Q_{T}(E)\W}=0$ for any $E\in\cal B(X)$
if and only if $\tr{\W\pi(s(x))T\pi(s(x)^{-1})}=0$ for $\alpha$-almost all
$x\in X$. Observing that $\pi(s(x))T\pi(s(x)^{-1})=\pi(g)T\pi(g^{-1})$ for
all $g\in G$ such that $p(g)=x$, this last condition is equivalent to  
$\tr{\W\pi(g)T\pi(g^{-1})}=0$ for $\mu$-almost all $g\in G$. Since the map
$g\mapsto \tr{\W\pi(g)T\pi(g^{-1})}$ is continuous, the lemma is proved.
\end{proof}

Let $G_1$ be the commutator subgroup of $G$, {\em i.e.} the subgroup of $G$
generated by the elements of the form $ghg^{-1}h^{-1}$, where $g,h\in G$, and 
assume that there is subspace $\ki$ of $\hi$ such that for all $g\in G_1$ and
$v\in\ki$, $\pi(g)v=c(g)v$ where $c$ is a character of $G_1$. 
Then the following result is obtained, compare with Th.~15 of~\cite{jmp}.
\begin{proposition}\label{prop2}
If $T$ is a state such that $T\hi\subset\ki$ and $\tr{T\pi(g)}\neq 0$ for
$\mu$-almost all $g\in G$, then $Q_{T}$ is  informationally
complete. 
\end{proposition}
\begin{proof}
  Let $\W$ be a trace class operator, 
and consider the decompositions of $T$ and $B$ as given in
  Section~\ref{s4}, {\em i.e.} $T=\sum_i \lambda_i |e_i\rangle\langle
  e_i|$ and $\W=\sum_k w_k |u_k\rangle\langle v_k| $ . Since
  $T\hi\subset\ki$, it follows that $\pi(g)e_i=c(g)e_i$ for all $g\in G_1$.
  Given $g\in G$, using the orthogonality relations~(\ref{ort1}), one has
\mfi 
& & \!\!\!\!\!\!\!\tr{T\pi(g)}\tr{\W\pi({g^{-1}})} \; = \;
\sum_{i,k} \lambda_i w_k \scal{e_i}{\pi(g)e_i}_\hi\scal{\pi(g)v_k}{u_k}_\hi \\
& &= \sum_{i,k} \lambda_i w_k \scal{c_{\pi(g)e_i,\pi(g)v_k}}{c_{e_i,u_k}}_{\hic} \\
&&= \;  \sum_{i,k} \ \lambda_i w_k\int_X \overline{c_ {\pi(g)e_i,\pi(g)v_k}(s(x))}
c_{e_i,u_k}(s(x))\,d\alpha(x)\\
& & =\;  \sum_{i,k} \ \lambda_i w_k\int_X
\scal{v_k}{\pi(s(x))\pi(s(x)^{-1}g^{-1}s(x)g)e_i}_\hi \scal{\pi(s(x))e_i}{u_k}_\hi d\alpha(x)\\
& & =\;  \sum_{i,k} \lambda_i w_k\int_X c(s(x)^{-1}g^{-1}s(x)g)
\scal{v_k}{\pi(s(x))e_i}_\hi \scal{\pi(s(x))e_i}{u_k}_\hi d\alpha(x)\\
& & =\;\int_X c(s(x)^{-1}g^{-1}s(x)g)
\tr{T\pi(s(x)^{-1})\W\pi(s(x))}\,d\alpha(x),
\mff
since $\sum_{i,k}  \lambda_iw_k  \scal{v_k}{\pi(s(x))e_i}\scal{\pi(s(x))e_i}{u_k}$ 
converges in $L^1(X,\alpha)$ to
  $\tr{T\pi(s(x)^{-1})\W\pi(s(x))}$, as shown in
  Section~\ref{s4}, and $c$ is bounded. Hence
$$
\tr{T\pi(g)}\tr{\W\pi({g^{-1}})} = \int_X c(s(x)^{-1}g^{-1}s(x)g)
\tr{T\pi(s(x)^{-1})\W\pi(s(x))}\,d\alpha(x)
$$
and, if $\tr{\W\pi(g)T\pi(g^{-1})}=0$ for all $g\in G$, then
$\tr{\W\pi({g^{-1}})}=0$ for $\mu$-almost all $g\in G$. 
On the other hand, if $\{e_n\}$ is a basis of $\hi$,
\mfi
\tr{\W\pi({g^{-1}})}  &=&  \sum_{n,p}\ip{e_n}{\W e_p}\ip{\pi(g)e_p}{e_n}\\
&=& \sum_{n,p}\ip{e_n}{\W e_p} (W_{e_p}e_n)(g),
\mff
where the double series converges in $\hic$. Since the set
$\{W_{e_p}e_n\}_{n,p}$ is orthonormal in $\hic$, the condition 
$\tr{\W\pi({g^{-1}})}=0$ for $\mu$-almost all $g\in G$ implies 
$\ip{e_n}{\W e_p}=0$ for all $n,p$, {\em i.e.} $\W=0$ and this proves that
$Q_{T}$ is informationally complete.
\end{proof}
\begin{remark}{\rm
  The condition that $G_1$ is represented by a character is automatically
  fulfilled (on the whole $\hi$) if 
$G_1$ is contained in the centre of $G$, whence
  $\pi|_{G_1}=\chi|_{G_1}$.}
\end{remark}
\begin{remark} {\rm 
  Suppose $G$ is a Lie group and let $\hi^{\omega}$ be the dense subspace of
  $\hi$ of analytic vectors for $\pi$. If $T$ has range in $\hi^{\omega}$,
  then the function $G\ni g\mapsto \tr{T\pi(g)}$ is analytic. This
  guarantees that $\tr{T\pi(g)}\neq 0$ for $\mu$-almost all $g\in G$.}
\end{remark}

\section{An example}\label{s6}
To discuss an example it is convenient to work with another
realization of the induced representation $(l,\hic)$.

Let $J$ be the unitary operator from $\hic$ onto $\lta$ given by
$$
(Jf)(x):=f(s(x)),\ \ \ x\in X.
$$
$J$ intertwines the imprimitivity system $(l,P)$ with
$(\tilde l,\tilde P)$, where
\mfi
(\tilde l(a) f)(x) & = & \chi(s(x)^{-1}as(a^{-1}. x))\, f(a^{-1}. x), \ \ \ a\in G,\\
(\tp(E) f)(x) & = & \chi_E(x) f(x),\ \ \  E\in\cal B(X),
\mff
with $f\in\lta$.

Given $u\in\hi$, if we compose  $W_u:\hi\to\hic$ of Section~\ref{s4} with $J$ 
we obtain an operator $\twu:\hi\to\lta$ explicitly given by
$$
(\twu v)(x) = \cuv(s(x)) = \ip{\pi(s(x))u}{v}_\hi.
$$
If $u$ is a unit vector, $\twu$ intertwines the operator measure $Q_u$, defined in Section~\ref{s4},
with the projection measure $\tp$, which is the minimal Neumark dilation of $Q_u$.

We denote by $\tmpi$ the image of $\mpi$ under the map $J$.
The analysis of $\mpi$, made in Section~\ref{s5}, can easily be translated into an analysis of $\tmpi$.

\

\iii
\item{}
$\tmpi$ is a closed subspace of $\lta$, invariant under $\tilde l$.
\item{}
Let $(e_n)_{n\geq 1}$ be a  basis of $\hi$. Then $\tmpi = \oplus_{p\geq 1}\ {\rm ran}\, \widetilde{W}_{e_p}$.

\item{}
For each $n\geq 1$, $(\tilde l_{{\rm ran}\, \widetilde{W}_{e_n}},{\rm
  ran}\, \widetilde{W}_{e_n})$ 
is equivalent
to the irreducible unitary representation $(\pi,\hi)$ of $G$.

\item{}
For each $n,p\geq 1$, we define
$$
f_{n,p}(x):= \widetilde{W}_{e_n}e_p.
$$
For each $n\geq 1$, $(f_{n,p})_{p\geq 1}$ is a basis of ${\rm ran}\, \widetilde{W}_{e_n}$,
and  $(f_{n,p})_{n,p\geq 1}$ is a  basis of $\tmpi$.
\fff

\subsection{The Heisenberg group}
We denote by $H^1$ the Heisenberg group. It is $\runo^3$ as a set  and we denote its elements
by $(t,q,p)$. The product rule is given by
$$
(t_1,q_1,p_1)(t_2,q_2,p_2)
=
(t_1+t_2+\frac{p_1q_2-q_1p_2}2,q_1+q_2,p_1+p_2).
$$
$H^1$ is a connected, simply connected,  
unimodular Lie group.
Its centre is $Z=\{(t,0,0)\,|\, t\in\runo\}$, and the quotient space $X=H^1/Z$
can be identified with $\runo^2$ (with respect to all relevant structures).
For the sake of convenience we choose the Haar measures $\mu, \mu_0$, and
$\alpha$ on $G$, $Z$, and $X$, respectively,  as $\frac 1{2\pi}dtdqdp$, $dt$,
and $\frac 1{2\pi}dqdp$.  The canonical projection $p:G\to X$ is the
coordinate projection $p((t,q,p))=(q,p)$, and we choose the natural, smooth
section $s((q,p))=(0,q,p), q,p\in\runo$.  With these choices the integral
formula of Section~\ref{s1}, which links together the measures $\mu, \mu_0$,
and $\alpha$ reads
$$
\int_{\runo^3} f(t,q,p)\,\frac {dtdqdp}{2\pi} =
\int_{\runo^2}\left( \int_{\runo}f((0,q,p)(t,0,0))\,dt
\right)\,\frac{dqdp}{2\pi},
$$
for all $f\in C_c(\runo^3)$, and is simply a consequence of Fubini's theorem.

Let $\hi$ be a complex separable infinite dimensional Hilbert space, and let
$(e_n)_{n\geq 1}$ be an orthonormal  basis of $\hi$.
There is a natural action of $H^1$ on $\hi$. Let $a,a^*$ denote the ladder operators associated
with the basis $(e_n)_{n\geq 1}$, and define
\mfi
Q&=& \frac 1{\sqrt 2}(a+a^*)\\
P&=& \frac 1{\sqrt{2}i}(a-a^*)
\mff
on their natural domains. Then
$$
(t,p,q)\mapsto e^{i(t+qP+pQ)}$$
is a {\em unitary irreducible} representation of $H^1$ on $\hi$. It is the {\em only} unitary irreducible
representation of $H^1$ whose central character is $t\mapsto e^{it}$, see for instance~\cite{Taylor86}.
It is {\em unitarily equivalent} to the representation of $H^1$
which acts on $L^2(\runo)$ as
$$
(\pi(t,q,p)\phi)(x) = e^{i(t+px+qp/2)}\phi(x+q),\ \ \ \phi\in L^2(\runo).
$$
We show that $(\pi,L^2(\runo))$ is a representation of $H^1$ that is square integrable
modulo the centre $Z$.
According to item 1 of section~\ref{s3}, it suffice to show that $c_{\phi,\phi}\circ s\in L^2(\runo^2)$
for some   $\phi\in L^2(\runo)$.
Explicitly 
$$
c_{\phi,\phi}(s(q,p))= \ip{\pi(s(q,p))\phi}{\phi}= 
e^{-i\frac{pq}2}\int e^{-ipx} \overline{\phi(x+q)}\phi(x)\, dx.
$$
Choose $\phi\in C_c(\runo)$, then, for any $q\in\runo$ 
$$
\left(x\mapsto \overline{\phi(x+q)}\phi(x) \right)\in\lyr\cap\ltr.
$$
Properties of the Fourier transform tell us that
$$
\left(p\mapsto   \int_{\runo}e^{-ipx}\overline{\phi(x+q)}\phi(x)\,dx \right)\in\ltr,\ \ \  q\in\runo.
$$
Thus we have, by the Plancherel theorem,
$$
\int_{\runo} \left|e^{-i\frac{pq}2}\int_{\runo}e^{-ipx}\overline{\phi(x+q)}\phi(x)\,dx\right|^2 dp
=
2\pi \int_{\runo} \left|\overline{\phi(x+q)}\phi(x)\right|^2dx,
$$
and, by the Fubini theorem,
$$
\int_{\runo} \left(2\pi \int_{\runo} |\overline{\phi(x+q)}\phi(x)|^2dx \right) dq= 2\pi\no{\phi}^4_{\ltr}.
$$
Tonelli's theorem tells us that the function  $c_{\phi,\phi}\circ s$ is in $L^2(\runo^2)$.
Moreover, recalling that $d\alpha=\frac{dq\,dp}{2\pi}$,
$$
\no{c_{\phi,\phi}\circ s}_{L^2(\runo^2,\alpha)}= \no{\phi}^2_{\ltr}.
$$
This shows that $\pi$ is square integrable modulo the centre and that its formal degree is $1$.
Since $\pi$ is the only irreducible representation of $H^1$ with the central
character $e^{it}$ and it is square integrable modulo the centre we conclude that
$$
\tmpi = L^2(\runo^2,\alpha),
$$
namely, that $(\tilde{l},L^2(\runo^2,\alpha))$ is an isotypic representation.
To exhibit this representation, let us observe that 
the map $\twu:\hi\to L^2(\runo^2,\alpha)$
is given by
$$
(\twu v)(x,y) = \ip{e^{i(xQ+yP)}u}{v}_\hi.
$$
The functions $f_{n,p}$, $p\geq 1$, which constitute a basis of ${\rm ran}\,\widetilde{W}_{e_n}$,
are 
$$
f_{n,p}(x,y) =  \sqrt{2\pi} \ip{e^{i(xQ+yP)}e_n}{e_p}_\hi.
$$
The  operator measure $Q_u$ is given by
$$
\ip{v}{Q_u(E)w}=
\frac 1{2\pi} \int_E\,\ip{v}{\pi(0,q,p)u}_\hi\ip{u}{\pi(0,q,p)^{-1}w}_\hi dqdp,
$$
which can be written as
$$
Q_u(E)= \frac 1\pi\int_E\, D_z|u\rangle\langle u|D_z^{-1}d\lambda(z),
$$
where $z=\frac{-q+ip}{\sqrt 2}$, $D_z = e^{it+za^*-\overline{z}a}$, and $\lambda$ is the Lebesgue measure on 
$\mathbb C$.
The action of $\tilde{l}$ on $L^2(\runo^2,\alpha)$ can directly be computed and we get
$$
(\tilde{l}(t,q,p)\tf)(x,y) = e^{i(t+\frac{xp-yq}2)}\tf(x-q,y-p).
$$

As a final remark we note that 
the commutator group of the Heisenberg group is contained in its center 
so that  if $T$ is a state such that 
$\tr{T\pi(g)}\neq 0$ for almost all $g\in H^1$,  then by Proposition~\ref{prop2}
the operator measure  $Q_T$ is informationally complete. This holds, in particular, if the
range of $T$ is contained in the subspace of $\hi$ of analytic vectors.

\end{document}